\documentclass[english]{revtex4}
\usepackage[T1]{fontenc}
\usepackage[latin9]{inputenc}
\usepackage{amsmath}
\usepackage{graphicx}
\usepackage{amssymb}
\usepackage{esint}
\usepackage{babel}

\begin{document}

\title{Ultrashort Focused Electromagnetic Pulses}

\author{Daniel an der Br\"ugge, Alexander Pukhov}

\affiliation{Institut f\"ur Theoretische Physik I,
             Heinrich-Heine-Universit{\"a}t D\"usseldorf, D-40225, Germany}

\date{\today}

\begin{abstract}
In this article we present a closed analytical description for few-cycle, focused electromagnetic pulses of arbitrary duration and carrier-envelope-phase (CEP). Because of
the vectorial character of light, not all thinkable one-dimensional (1D)
shapes for the transverse electric field or vector potential can be realized as finite energy
three-dimensional (3D) structures. We cope with this problem by
using a second potential, which is defined as a primitive to the vector
potential. This allows to construct fully consistent 3D wave-packet solutions for the Maxwell equations, given a solution of the scalar wave equation. The wave equation is solved for ultrashort,
Gaussian and related pulses in paraxial approximation. The solution is given in a closed and numerically convenient form, based on the complex error function. All results undergo thorough 
numerical testing, validating their correctness and accuracy. A reliable and accurate representation of few-cycle pulses is e.g. crucial for analytical and numerical theory of vacuum particle acceleration.
\end{abstract}
\maketitle
\section{Introduction}
\label{sec:intro}
Recent developments in laser technology \citep{fat lasers} resulted in
ultrashort electromagnetic pulses, which may 
contain only a few optical cycles 
and can be focused down to a single wavelength leading to the so-called
$\lambda^3$-regime. In addition, there was a tremendous
experimental progress in carrier-envelope-phase (CEP) 
control of these ultrashort laser pulses allowing to synthesize
almost arbitrary pulse shapes \citep{CEP control Jones,CEP
control Xu,CEP relevance}. Applications for these well controlled
laser pulses range from coherent attosecond control
\citep{attoscience} to high-gradient electron
acceleration \citep{salamin, salamin RP, anupam, canada} and generation of ultrashort
coherent X-ray flashes \citep{Naumova, baevaRPC, Tsakiris, NatureAtto}. At the same time, numerical studies demonstrate 
the importance of correct analytical description of laser pulses in vacuum \citep{salamin,salamin RP}. It was shown, 
that even weakly inaccurate solutions of Maxwell equations can lead to
largely erroneous results when applied blindly to, e.g., direct particle acceleration by the laser fields. Clearly, there is a demand for accurate analytical description of these pulses. 

As in the case of strongly focused pulses  \citep{paraxial}, the vectorial character of light becomes crucial for few-cycle pulses. There is a significant interdependence between the pulse shape and the
polarization that requires careful analysis.
We demonstrate that not all field structures conceivable in 1D models can be realized as finite energy, localized
3D wave packets. Commonly used approximations are consistent
for a certain choice of the carrier envelope phase (CEP) only. When trying to construct
laser pulses starting from a given shape for the transverse field component, 
one easily ends up with a pulse inconsistent in the 3D geometry. Our
method to construct consistent 3D electromagnetic structures is valid for arbitrary CEP cases.

Following the work of Porras \citep{porras,porras2}, our approach to the wave equation uses the analytic signal \citep{Bracewell}. We also consider the particularly interesting case of a radially polarized laser pulse \citep{RP PRL,RP Quabis}. 
Because of its strong and purely longitudinal field component on-axis, the radially polarized pulse may become an
important tool for electron acceleration \citep{salamin RP,anupam,canada}. We provide the proper analytical solutions in a simple manner, which is particular convenient for use in numerical simulations.

Finally, we let the solutions undergo some accurate numerical tests. Any significant errors in the solution would show up while they are propagated by the field solver. Compared to more conventional approximations, the new pulse description decreases electromagnetic artefacts drastically at the pulse initialization stage, and the self-consistent development of the pulse fields agrees with the analytical description in cases, where more conventional approximations fail.

\section{Second Potential Representation}
\label{sec:SecPot}

An electromagnetic pulse can be represented by its four-potential
$A^{\alpha}=(\phi,\mathbf{A})$, where each component satisfies the vacuum wave
equation $\square A^{\alpha}=0$. We use the Lorenz gauge $\partial_{\alpha}A^{\alpha}$=0
and further set the scalar potential to zero $\phi=0$, which can
be done in vacuum. Then, the fields are written as $\mathbf{E}=-c^{-1}\partial_{t}\mathbf{A}$
and $\mathbf{B}=\nabla\times\mathbf{A}$. We are interested in finite energy
pulse-like structures, so that $\mathbf{A}$  is required
to be a localized function: $|A|\rightarrow0$ for $r\rightarrow\infty$.
It is easy to see that the pulse potential is uniquely defined 
now, since each change of $\mathbf{A}$ generates measurable
electric or magnetic fields. Because of $\phi=0$, the Lorenz gauge
coincides with the Coulomb gauge in vacuum 
$\nabla\cdot\mathbf{A}=0$.

\begin{figure}
\includegraphics[width=1\columnwidth]{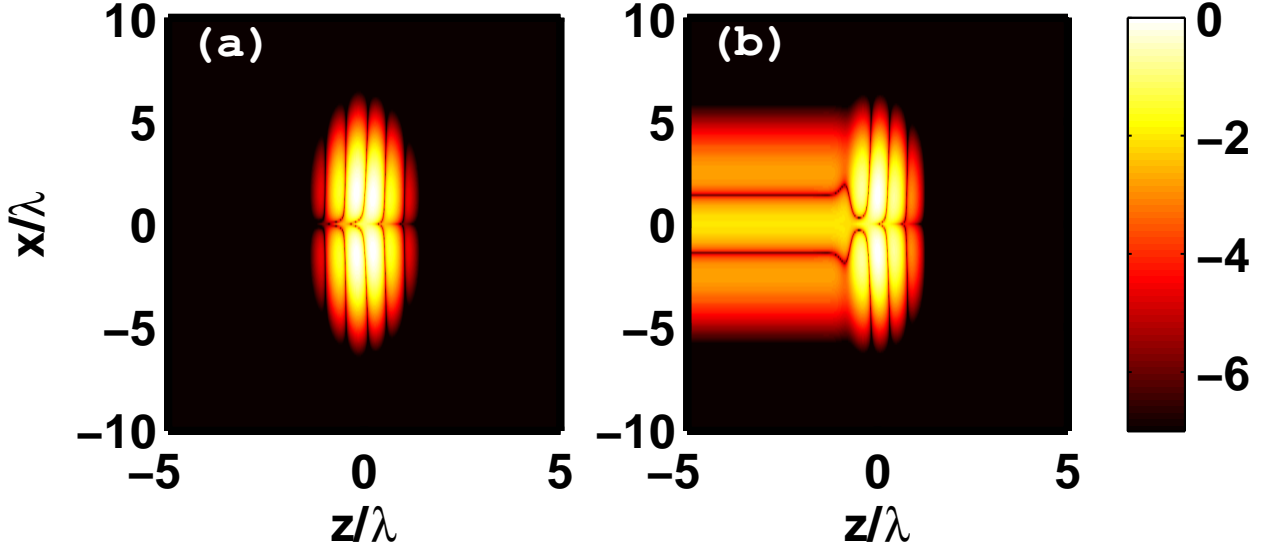}\caption{\label{fig:sin_vs_cos}Logarithm of the square transverse magnetic
field $\log_{10}\left((B_{y}/B_{0})^{2}\right)$ in the $y=0$ plane
resulting from an ultrashort, linearly polarized, (a) sine- and (b)
cosine-phased Gaussian potential. The pulse duration is $c\tau=0.5\,\lambda$
and focal spot width $\sigma=2\,\lambda$.}
\end{figure}

In laser physics, it is common to choose an analytical solution to
the wave equation for the main, i.e. transverse
components of the pulse. Then the longitudinal component can be determined: 
 $A_{z}=\int_{z}^{\infty}\nabla_{\perp}\cdot\mathbf{A}_{\perp}dz$.
However, this integral can yield a non-vanishing longitudinal potential
component far from the actual pulse region, since commonly used solutions
to the wave equation do not satisfy the condition 

\begin{equation}
\int_{-\infty}^{\infty}\nabla_{\perp}\cdot\mathbf{A}_{\perp}dz\overset{!}{=}0\label{eq:finiteness_condition}
\end{equation}

\noindent for ultrashort pulses. Note that this component is not meaningless
but will cause non-zero longitudinal electric and transverse magnetic
fields, an example of which is shown in Fig.~\ref{fig:sin_vs_cos}b.
These fields have a small amplitude of the order $\mathcal{O}\left((c/\omega\sigma)^{2}\right)$,
but since they extend infinitely along the beam axis, they contain an infinite amount of energy.

To get a realistic finite energy pulse, our choice of the transverse
vector potential is restricted by Eq.~\eqref{eq:finiteness_condition}.
This is a fundamental difference to the 1D case, where
such a restriction on the wave form does not exist. Before
the consequences of this restriction are discussed in detail, let us introduce the
second potential $\bf{\Psi}$, which enables us to describe a reasonable set of
realistic pulse structures in a more convenient way:

\begin{equation}
\mathbf{A}=\nabla\times{\bf \Psi}\label{eq:second_potential}\end{equation}

\noindent Of course, each component of $\bf \Psi$ has to satisfy the wave
equation $\Delta\Psi_{i}-\frac{1}{c^{2}}\partial_{t}^{2}\Psi_{i}=0\label{eq:wave_Psi}$. 
Then, the wave equation for $\mathbf{A}$ and the Coulomb gauge
readily follow from  Eq.~\eqref{eq:second_potential} and there are no restrictions like
\eqref{eq:finiteness_condition} 
on the choice of the second potential components.

For a large class of laser pulses, it is convenient to choose
$\hat{\mathbf{e}}_{z}\cdot{\bf \Psi}=0$. 
Then Eq.~\eqref{eq:second_potential} becomes
$A_{z}=-\nabla_{\perp}\cdot\left(\hat{\mathbf{e}}_{z}\times{\bf \Psi}\right)$
and $\mathbf{A}_{\perp}=\hat{\mathbf{e}}_{z}\times\partial_{z}{\bf \Psi_{}}$.
In the near-monochromatic case $\partial_{z}\sim-i\omega/c$ we get
$\hat{\mathbf{e}}_{z}\times{\bf \Psi}\approx i c \mathbf{A}_{\perp}/\omega$,
hence in the long pulse limit the transverse second potential (TSP) 
$\hat{\mathbf{e}}_{z}\times{\bf{\Psi}}$ is, except for a constant factor,
identical with the transverse components of the vector potential.

Let us point out that each electromagnetic pulse in vacuum
can be represented by the TSP. This is easily seen, since the second
potential can be obtained by taking the integral $\hat{\mathbf{e}}_{z}\times\bf{\Psi}=\int_{z}^{\infty}\mathbf{A}_{\perp}dz'$
from an arbitrary vector potential in Coulomb gauge. To make the representation unique, we
require the condition $|{\bf{\Psi}}|\rightarrow0$ for $|\mathbf{r}|\rightarrow\infty$
in the half-space $x>0$, so that $\bf{\Psi}_{\perp}$ is unambiguously
given by the just mentioned integral. Then, for a vast class of laser
pulses including all linearly, circularly and radially polarized modes,
${\bf{\Psi}}$ will vanish at infinity in all directions. To understand
this, we write the integral condition \eqref{eq:finiteness_condition}
in terms of the TSP:

\begin{equation}
\nabla_{\perp}\cdot\left(\hat{\mathbf{e}}_{z}\times\left.\Psi\right|_{z=-\infty}\right)=0\label{eq:finiteness_psi}
\end{equation}

In general, it possesses non-trivial solutions, corresponding to structures,
where the fields produced by ${\bf{\Psi}}$ at $ z \rightarrow-\infty$ vanish,
but ${\bf{\Psi}}$ itself does not. However, we look at important special
cases. For linear polarization ($\Psi_{x}=0$), Eq.~\eqref{eq:finiteness_psi}
has obviously none but the trivial solution. Thus, all finite energy,
linearly polarized pulses can be represented by a localized TSP.
The same is true for circular polarization, which we define by $\Psi_{y}=i\Psi_{x}$,
and the components are assumed to be analytical functions. Eq.~\eqref{eq:finiteness_psi}
then only has solutions of the type $\Psi(x+iy)$, so there is no non-trivial
solution that fulfills the boundary condition. Another interesting
structure is the radially polarized pulse ($\hat{\mathbf{e}}_{z}\times\mathbf{\Psi}=f(r_{\perp},z,t)\mathbf{r}_{\perp}$),
and again there is no non-trivial finite energy solution to \eqref{eq:finiteness_psi}
obeying its symmetry. 

Summarizing, for each realistic linearly, circularly
or radially polarized pulse, the condition $\hat{\mathbf{e}}_{z}\times\left.\bf{\Psi}\right|_{z=-\infty}=\int_{-\infty}^{\infty}\mathbf{A}_{\perp}dz\overset{!}{=}0$
on the transverse vector potential must be
fulfilled. Commonly used analytical wave packet solutions
strictly fulfill this condition only for a certain choice of the CEP.
Take, for instance, a pulse with a Gaussian longitudinal
profile, $A_{y}(x=0,\, y=0,\, z,\, t=0)=a_{0}\exp\left[-(z/c\tau)^{2}\right]\cos(2\pi z/\lambda)$,
then $\left|\Psi_{y}(z=-\infty)\right|/(a_{0}\lambda)=\pi^{-1/2}\overline{\tau}\,\exp\left(-\overline{\tau}^{2}\right)\neq0$, meaning that some field components would extend to infinity as in Fig.~\ref{fig:sin_vs_cos}b.
Unlike the cosine-phased Gaussian, the sine-phased potential may in principle be assigned to $A_y$, as the integral vanishes here (see also Fig.~\ref{fig:sin_vs_cos}a). Assigning an exact Gaussian profile to the transverse electric field instead does not help. Indeed, by calculating the second potential of such a pulse it can be proven
that this structure does not exist for any phase. 

Other than for linearly, circularly or radially polarized pulses, for
the azimuthally polarized pulse $\hat{\mathbf{e}}_{z}\times\mathbf{\Psi}=f(r_{\perp},z,t)\hat{\mathbf{e}}_{\varphi}$
the TSP representation  not have to vanish as $z\rightarrow-\infty$.
Notice that this pulse does not produce any longitudinal field at
all. Anyway, it can neatly be represented by the longitudinal component $\hat{\mathbf{e}}_{z}\cdot{\bf \Psi}$
instead of the transverse components of the second potential.

\section{Scalar Wave Equation}
\label{sec:ScalWav} 

Now we come to the solution of the scalar wave equation on $\bf \Psi$.
The Fourier transform in time and the transverse
directions yields $\partial_{z}^{2}\tilde{\Psi}=-((\omega/c)^{2}-k_{\perp}^{2})\tilde{\Psi}$,
with the solution 
\begin{equation}
\tilde{\Psi}(x,\mathbf{k}_{\perp},\omega)=\left.\tilde{\Psi}\right|_{z=0}\exp\left(-iz\sqrt{\left(\frac{\omega}{c}\right)^{2}-k_{\perp}^{2}}\right)\label{eq:solution_wave}
\end{equation}
Next, the focal spot profile $\tilde{\Psi}(z=0)$ has to be chosen
in a physically reasonable way. Besides the very common linearly polarized Gaussian mode $\textrm{TEM}_{00}$, we also treat the very interesting radially polarised Hermite-Gaussian mode $\textrm{TM}_{01}$. Further we will tackle the 2D solutions,
which differ in some factors from the 3D ones. Knowing the scalar
solution for the $\textrm{TEM}_{00}$ mode, we can easily construct circularly polarized Gaussian
pulses by setting $\Psi_{y}=i\Psi_{x}$, and moreover azimuthally
polarized pulses by assigning the same term to the
longitudinal component $\hat{\mathbf{e}}_{z}\cdot\bf{\Psi}$ of the
second potential.

Let $\psi(t)$ be the pulse time dependence at the center of the focal
spot and $\tilde{\psi}(\omega)$ its Fourier transform. The focal
spot size may depend on the frequency:

\begin{equation}
\left(\hat{\mathbf{e}}_{z}\times\left.\tilde{\bf{\Psi}}\right|_{z=0}\right)=\tilde{\psi}(\omega)\,\exp\left[-\left(\frac{r_{\perp}}{\sigma\left(\omega\right)}\right)^{2}\right]\begin{cases}
\mathbf{r}_{\perp} & \textrm{TM}_{01}\\
\hat{\mathbf{e}}_{x} & \textrm{TEM}_{00}\end{cases}\label{eq:spot_profile}\end{equation}

Now, the transverse direction Fourier transform of \eqref{eq:spot_profile}
is inserted into the solution of the wave equation \eqref{eq:solution_wave}.
Since we consider pulses propagating mainly in one direction ($c^{2}k_{\perp}^{2}\ll\omega^{2}$),
we expand the square root in a Taylor series and neglect fourth order
terms: $\sqrt{\left(\omega/c\right)^{2}-k_{\perp}^{2}}\approx\omega/c-ck_{\perp}^{2}/(2\omega)$,
performing the paraxial approximation. If now $\sigma\propto1/\sqrt{\omega}$
is assumed, so that the Rayleigh length $z_{Rl}=\omega\sigma^{2}/2c$
is constant for all frequencies, we are able to carry out the inverse
Fourier transform analytically and obtain the solution 
\begin{equation}
\hat{\mathbf{e}}_{z}\times\mathbf{\Psi} \left(\mathbf{r},t\right)=\left(\frac{z_{Rl}}{q}\right)^{g}\,\psi\left(t-\frac{z}{c}-\frac{r_{\perp}^{2}}{2cq}\right)\begin{cases}
\mathbf{r}_{\perp} & \textrm{TM}_{01}\\
\hat{\mathbf{e}}_{x} & \textrm{TEM}_{00}\end{cases}\label{eq:complete_solution}
\end{equation}

\begin{figure}
\includegraphics[width=1\columnwidth]{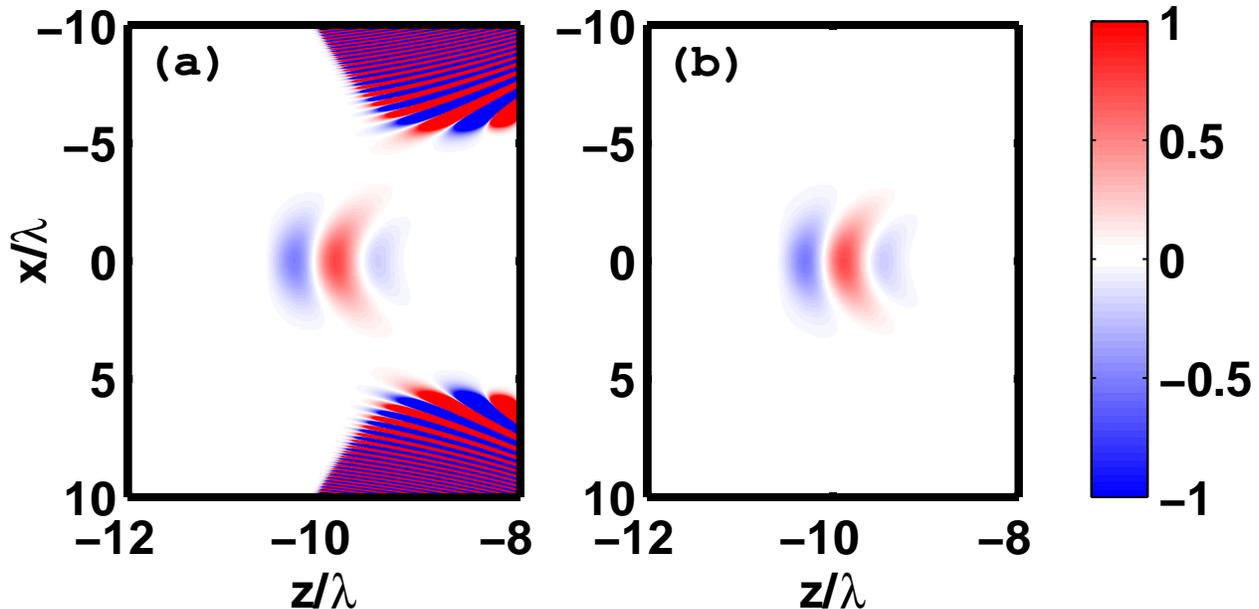}

\caption{\label{fig:divergent_vs_analytic}Direct comparison of a two dimensional
cut through the complete short-pulse solution \eqref{eq:complete_solution}
using (a) the {}``naive'' choice $\psi(t')=\exp\left(-t'^{2}/\tau^{2}+i(\omega_{0}t'+\phi)\right)$
and (b) the analytic signal Eq.~\eqref{eq:analytic_Gauss}. Pulse
parameters are: $c\tau=0.5\,\lambda$, $\sigma=2\,\lambda$, $ct=-10\,\lambda$
(before focus)}

\end{figure}

Here $g=1$ for a linear ($g=0.5$ in 2D) and $g=2$ for a RP ($g=1.5$
in 2D) laser and $q=z+iz_{Rl}$ is the confocal parameter. $\psi$
is a complex representation of the time dependence of the pulse. Note,
that since $t'=t-z/c-r_{\perp}^{2}/2cq$, it is generally a complex
number. Choosing naively $\psi(t')=\exp\left(-t'^{2}/\tau^{2}+i(\omega_{0}t'+\phi)\right)$
yields a solution diverging for big $r_{\perp}$ as $\mathcal{O}(\exp(r_{\perp}^{4}))$.
Instead the analytic signal \citep{Bracewell} should be used, as
suggested by Porras \citep{porras}. The analytic signal is the complex
representation of a real signal without negative frequency components.
The analytic signal representation of the Gaussian pulse $g(t')=\exp(-t'^{2}/\tau^{2})\cos(\omega_{0}t+\phi)$
is calculated to be 

\begin{equation}
\psi\left(t'\right)=\psi_{0}\frac{e^{-\overline{\tau}^{2}}}{2}\left(e^{i\phi}w\left(\frac{t'}{\tau}-i\overline{\tau}\right)+e^{-i\phi}w\left(\frac{t'}{\tau}+i\overline{\tau}\right)\right)\label{eq:analytic_Gauss}
\end{equation}

\noindent wherein $\overline{\tau}=\omega_{0}t/2$ and $w\left(z\right)=\textrm{exp}(-z^{2})\textrm{erfc}(-iz)$
is the complex error (or Faddeeva) function \citep{Abramowitz}. The analytic signal of the Gaussian pulse
was also calculated in \citep{porras}, but Eq.~(27) from \citep{porras}
disagrees with \eqref{eq:analytic_Gauss} for $\textrm{Im}(t')\neq0$.
Analytical and numerical tests show, that \eqref{eq:analytic_Gauss}
is the correct solution. To illustrate the meaning of \eqref{eq:analytic_Gauss}
and its difference to the naive choice $\psi(t')=\exp\left(-t'^{2}/\tau^{2}+i(\omega_{0}t'+\phi)\right)$,
consider Fig.~\ref{fig:divergent_vs_analytic}. In the region near
the optical axis, $\textrm{Im}(t')$ is small and the solutions agree
quite well. However, for bigger $r_{\perp}$ the naive solution (b)
diverges, while the analytic signal shows a proper beam-like behaviour
and vanishes.

\section{Numerical Testing}
\label{sec:Numerical}

The equations \eqref{eq:second_potential}, \eqref{eq:complete_solution}
and \eqref{eq:analytic_Gauss} together form the key to accurate analytical
and numerical representation of ultrashort few- and even single-cycle
electromagnetic pulses. One important application of them is the use
in numerical simulations and we will conclude this paper by showing
their superiority to more conventional representations for this application, thereby checking the correctness and the
accuracy of the analytical results obtained so far. 

We use the particle in cell (PIC) code VLPL \citep{vlpl}. To begin
with, the fields are initialised inside the VLPL simulation grid.
Then they are propagated using a standard algorithm on the Yee-mesh.
Finally, it is verified if the numerically propagated pulse still
agrees with the analytical term, overall or in some key parameters,
and furthermore, if unphysical static fields remain at the place,
where the pulse was initialized.

\begin{figure}
\includegraphics[width=1\columnwidth]{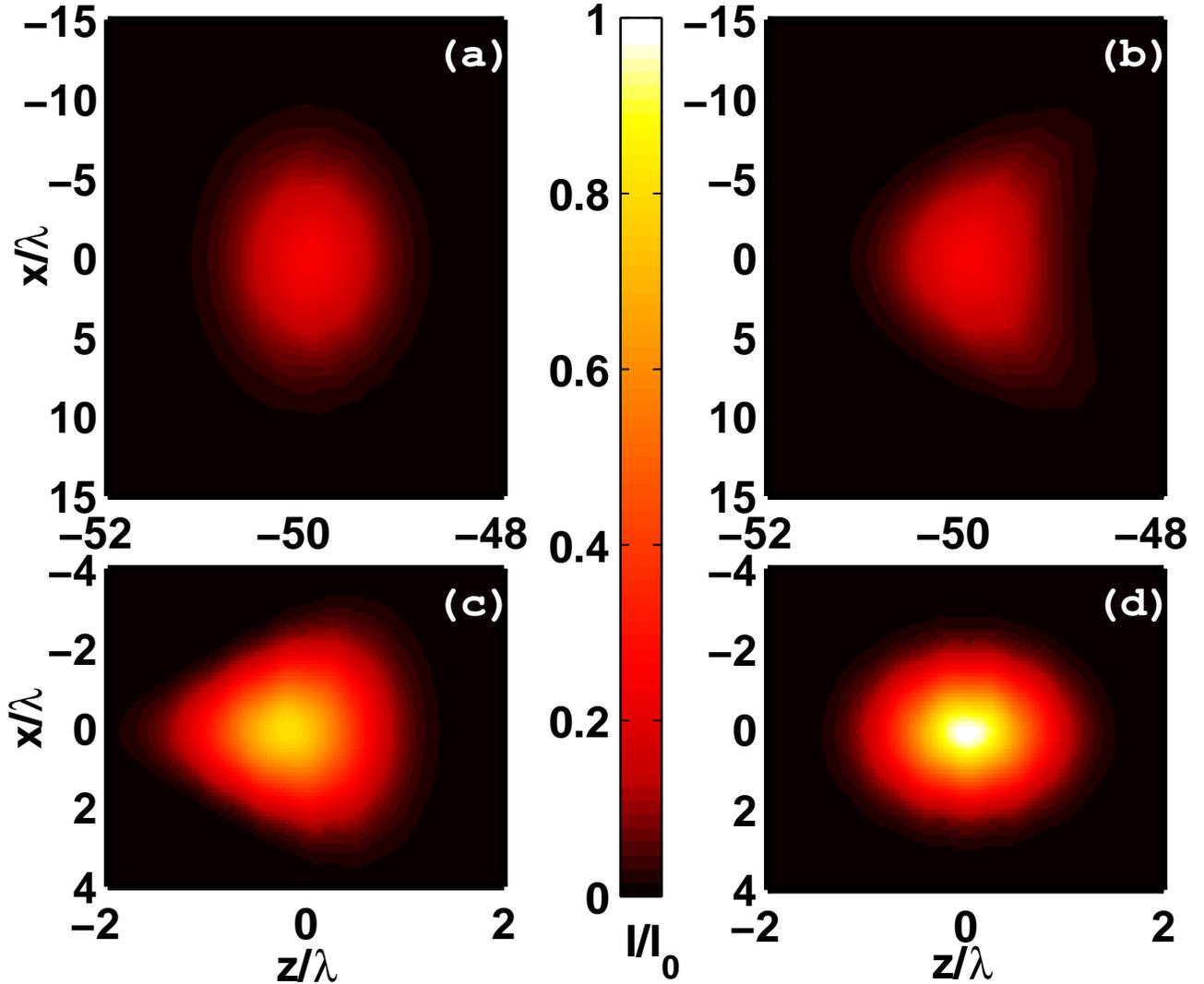}

\caption{\label{fig:focusing}Focusing properties of different approximations
of the wave equation, evaluated in a 2D version of the PIC code VLPL.
(a) and (b) show the intensity distribution as it is initialised inside
the code and (c) and (d) show the propagated solution at the focal
spot. (a) and (c) use the CW paraxial solution multiplied with a temporal
profile, (b) and (d) the correct short pulse solution. The circularly
polarized laser pulses are Gaussian both in space and time with duration
$c\tau=\lambda$ and width $\sigma=2\lambda$.}

\end{figure}

First we check the correctness of Eqs.~\eqref{eq:complete_solution}
and \eqref{eq:analytic_Gauss}, proving their superiority to a conventional
representation often used for numerical simulations. The ``conventional'', or separable form is a simple product of a monochromatic, transversely Gaussian beam with a Gaussian temporal profile. Circularly polarized
pulses are used, so that the shape can well be seen in the intensity
plots, with a duration of $c\tau=\lambda$ and a focal spot width
of $\sigma=2\lambda$. The pulses are focused over a distance of $50\lambda$
inside the simulation. In Fig.~\ref{fig:focusing} the initial
condition and the numerically propagated solution at the focal spot
are shown. While the product approximation shows strong asymmetric
deformation in the focus and the focal field does not reach its specified
value of $a_{0}=1$, the proper short pulse solution is nearly perfectly
symmetric in the focus and reaches the desired maximum. The presented
solution Eqs.~\eqref{eq:complete_solution} and \eqref{eq:analytic_Gauss}
is clearly superior to the simple product approach.

Now we come to the second potential representation. The alternative
to its use is to assign arbitrary wave equation solutions to the transverse
components of the vector potential and then make some kind of approximation
for the longitudinal part. The simplest possibility is to fully neglect
the longitudinal field component, but for strongly focused pulses
a somewhat more reasonable approximation can be reached by choosing
$A_{z}=(c/i\omega_{0})\nabla_{\perp}\cdot\mathbf{A}_{\perp}$, what
we will call the quasi-monochromatic approximation, because it relies
on $\partial_{z}\sim-i\omega_{0}/c$. Using one of these approximations,
the initial pulse structure has finite energy, and because of the
energy conserving property of the field propagator algorithm, the
pulse will be forced to self-organize into a consistent structure
of finite extension. While this happens, ``virtual charges'' are left behind, unphysical static fields, which may make the concerned regions in the simulation domain unusable for further computations. We want to see, if this problem
can be cured by the use of the second potential representation.

\begin{figure}
\includegraphics[width=1\columnwidth]{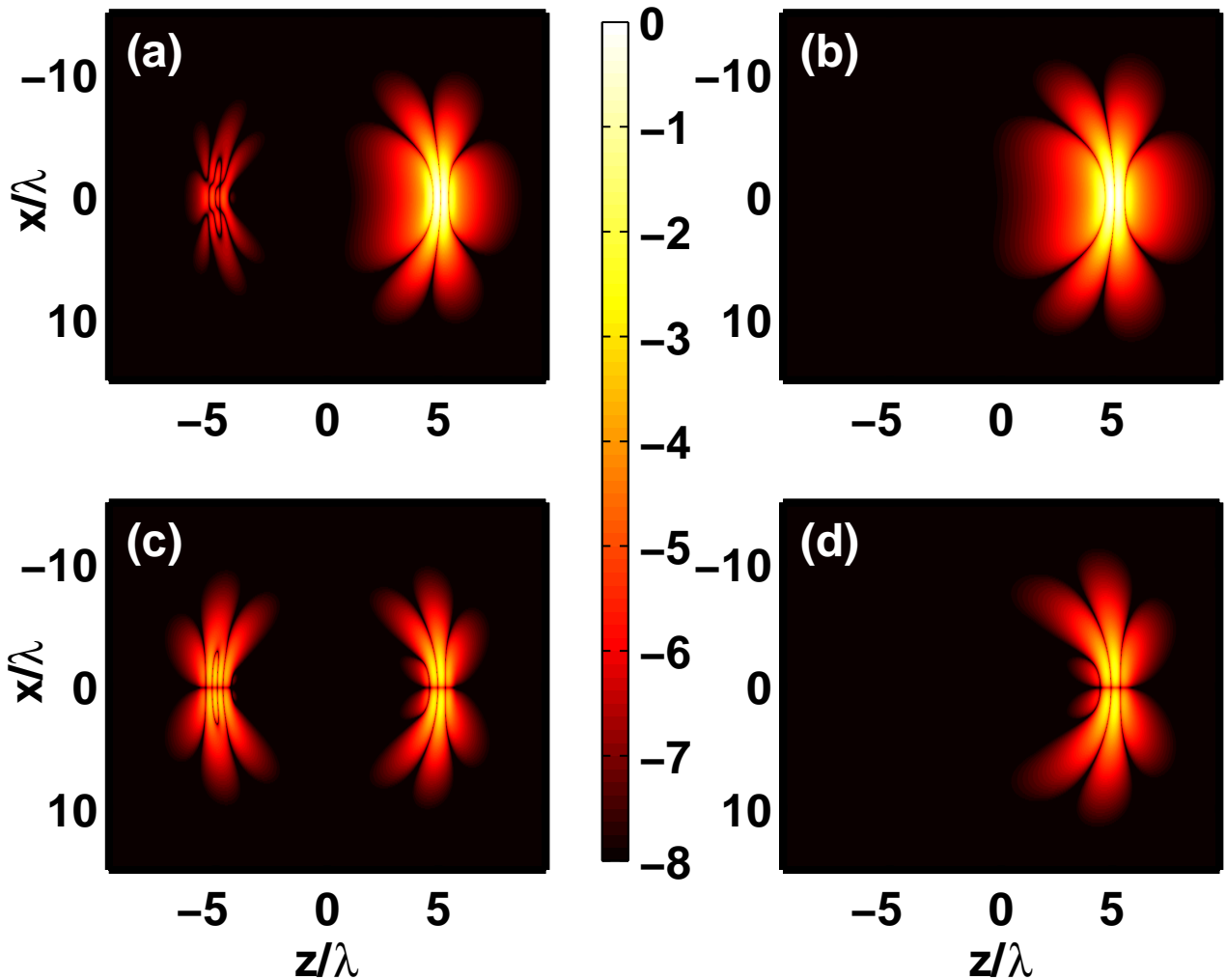}\caption{\label{fig:gauss_conventional_vs_tsp_propagated}Logarithm of the
transverse ((a) and (b)) and longitudinal ((c) and (d)) square electric
field $\log_{10}\left((E_{i}/E_{0})^{2}\right),\; i\in\{x,z\}$ after
a propagation time $ct=10\,\lambda$. (a) and (c) depict the conventional,
(b) and (d) the TSP Gaussian. Pulse parameters are $c\tau=0.3\,\lambda$
and $z_{Rl}=4\pi^{2}\,\lambda$. For the second potential, a corrected
frequency $\omega_{0}'=\omega_{0}-\sqrt{2}/\tau$ was used to take
account for the frequency shift caused by the additional derivative.}
\end{figure}

When employing the second potential representation for an ultrashort
pulse, the additional derivative will slightly alter the pulse shape.
As shown before, this is inevitable, since a Gaussian shaped linearly
polarized vector potential can exist as an independent structure in
vacuum only if it is sine-phased. To make the TSP represented pulse
comparable to the conventional one, it is necessary to take care for
the frequency shift caused by the $z$-derivative, which
can be estimated as $\omega_{\textrm{eff}}=\omega_{0}+\sqrt{2}/\tau$.

Fig.~\ref{fig:gauss_conventional_vs_tsp_propagated} shows the
pulses after a short propagation distance in the PIC simulation box.
Firstly one observes that, despite of the very short duration, the
moving pulse structures (right half of the images) appear very similar
in the conventional and the TSP version. Secondly one notices, that
the conventional pulse leaves behind a significant amount of virtual
charge fields in the initialisation region (left half of the images),
having both a longitudinal and a transverse component. This undesired
phenomenon can neatly be suppressed by the use of the second potential,
seen in the figure.


The last test we want to present in this paper concerns the longitudinal field component on the optical axis of a radially polarized pulse, which is of particular interest for vacuum electron acceleration \citep{salamin RP,anupam,canada}. The pulse length used was $c\tau=0.5\lambda$, and the focal spot size $\sigma=2\lambda$. Again, the conventional approach corresponds to a near-monochromatic approximation
$A_{x}=i\nabla_{\perp}\cdot\mathbf{A}_{\perp}/\omega$, so as to generate
a finite pulse structure from the given transverse field components. The phase $\phi$ in Eq.~\eqref{eq:analytic_Gauss}
was chosen as $\phi=0$ for the second potential representation and
as $\phi=0.5\,\pi$ for the conventional representation, so that the
pulses are actually comparable. Initially, the longitudinal field nearly agrees for both representations, since the first term of Eq.~\eqref{eq:analytic_Gauss}, which is the dominating one, is the same in both cases. 

\begin{figure}
\includegraphics[width=1\columnwidth]{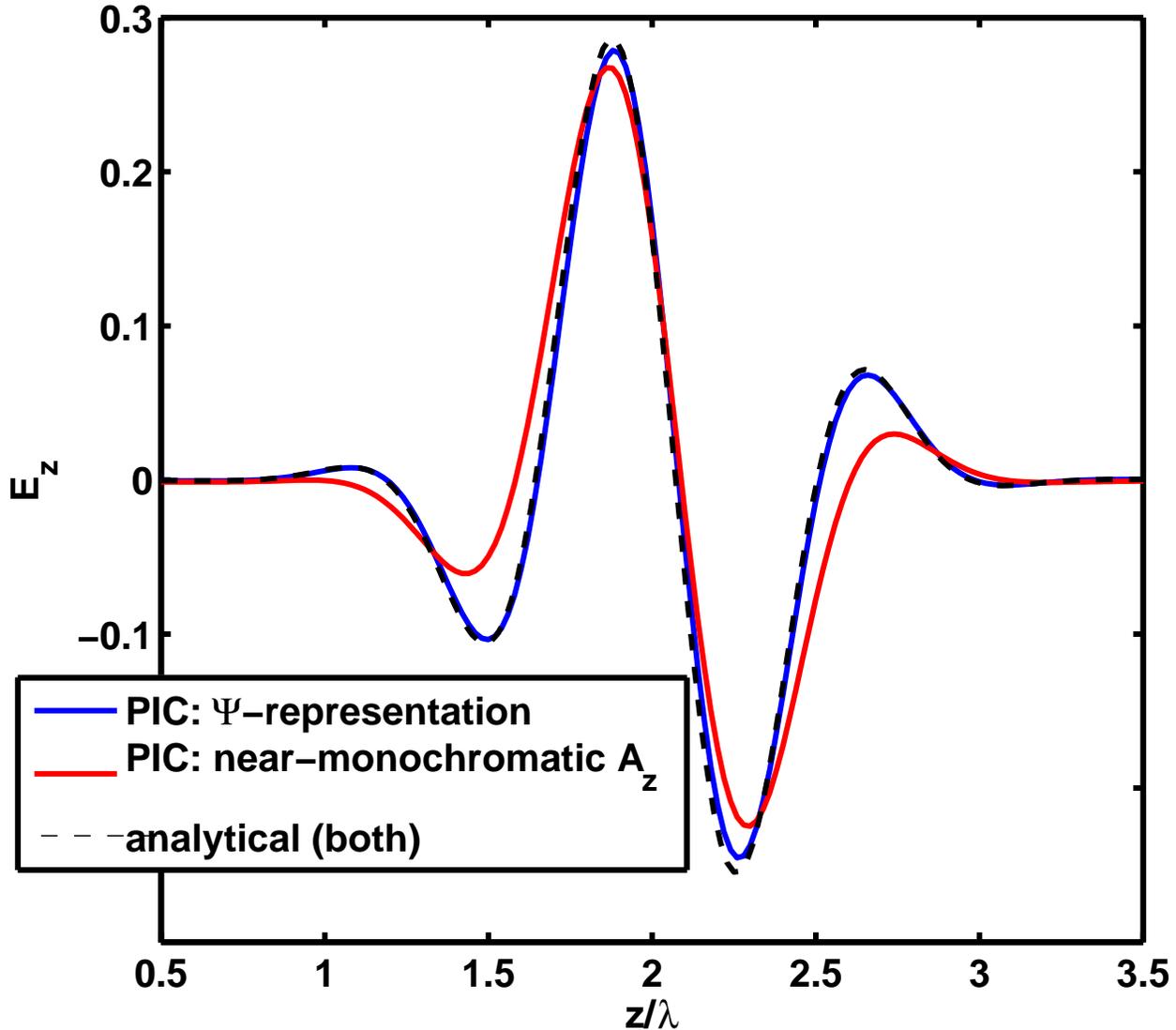}

\caption{\label{fig:long_field_on_axis}The longitudinal pulse field in PIC
at the time $c(t-t_{0})=4\lambda$ after initialization, compared
to the analytic solution. Once the TSP was used for initialization
(blue line) and once the near-monochromatic approximation for $A_{z}$.
The differences in the analytical representations are too small to
be seen in the diagram, so that both are represented by one curve
(dashed black line). Pulse parameters are: $c\tau=0.5\lambda$, $\sigma=2\lambda$,}

\end{figure}

In figure~\ref{fig:long_field_on_axis} we present the longitudinal
field of the pulse after it has left its ``virtual charges'' behind. The fields initialized using the conventional representation already differ significantly from the analytical description,
whereas the TSP represented fields agree almost perfectly. This will be crucial e.g. when PIC simulations
are to be compared with other analytical or semi-numerical calculations,
where the field is inserted analytically and is not self-consistently propagated.
Without an exact and reliable pulse representation, such a comparison
is hardly possible.

\section{Conclusion}
\label{sec:Conclusion}

In this paper we have given a comprehensive guide to the mathematical
representation of ultrashort, Gaussian and related, electromagnetic
pulses. The wave equation for an ultrashort Gaussian pulse has been
solved in paraxial approximation. Further it has been shown, that
the vectorial character of light has a stringent influence on its
field structure for ultrashort pulses. To pay regard to this, ultrashort
pulses should be represented by their second potential, using the
transverse components for linear, circular or radial polarization
and the longitudinal component for azimuthal polarization. Numerical tests proove that the solutions flawlessly work in the regime of ultrashort, moderately strong ($\sigma\gtrsim\lambda$) focused pulses.

\section*{Acknowledgements}
\label{sec:Ack}
This work has been supported by GRK1203 (DFG, Germany).


\begin{thebibliography}{10}
\bibitem{fat lasers}G.~A.~Mourou, et al. Plasma~Phys.~Control.~Fusion~\textbf{49}
B667-B675 (2007).

\bibitem{CEP control Xu}L.~Xu, Ch.~Spielmann, A.~Poppe, T.~Brabec,
F.~Krausz, and T.~W.~H\"ansch, Opt.~Lett.\textbf{~21}(24), 2008
(1996).

\bibitem{CEP control Jones}David~J.~Jones, et al. Science\textbf{~288},
635 (2000).

\bibitem{CEP relevance}G.~G.~Paulus, et al. Nature~(London)\textbf{~414},
182-184 (2001).

\bibitem{attoscience}P.~B.~Corkum, F.~Krausz, Nature~Phys.~\textbf{3}, 381 (2007)
\bibitem{salamin}Y.~I.~Salamin and C.~H.~Keitel, Phys.~Rev.~Lett.~\textbf{88}(9), 095005 (2002).

\bibitem{salamin RP}Y.~I.~Salamin, Phys.~Rev.~A~\textbf{73}, 043402 (2006).

\bibitem{canada}C.~Varin, M.~Piche, and M.~A.~Porras, Phys.~Rev.~E~\textbf{71}, 026603 (2005).

\bibitem{anupam}A.~Karmakar and A.~Pukhov, Laser~and~Particle~Beams~\textbf{25},
371 (2007).
\bibitem{Naumova}N.~M.~Naumova, J.~A.~Nees, B.~Hou, G.~A.~Mourou and I.~V.~Sokolov, Opt.~Lett.~\textbf{29}(7), 778 (2004).
\bibitem{baevaRPC}T.~Baeva, S.~Gordienko and A.~Pukhov, Phys.~Rev.~E~\textbf{74}, 065401(R) (2006).
\bibitem{Tsakiris} G.~D.~Tsakiris, K.~Eidmann, J.~Meyer-ter-Vehn, and F.~Krausz, New~J.~Phys. {\bf 8}, 19 (2006).
\bibitem{NatureAtto}  A.~Pukhov, Nature Physics, {\bf 2}, 439 (2006).
\bibitem{paraxial}M.~Lax, W.~H.~Louissell, W.~B.~McKnight, Phys.~Rev.~A~\textbf{11}(4),
1365 (1975).
\bibitem{porras}M.~A.~Porras, Phys.~Rev.~E~\textbf{58}(1), 1086
(1998).
\bibitem{porras2}M.~A.~Porras, Phys.~Rev.~E~\textbf{65}, 026606 (2002).

\bibitem{Bracewell}R.~Bracewell, \emph{The Fourier Transform and
Its Applications}, (McGraw-Hill, 2nd ed. 1986).
\bibitem{RP Quabis}S.~Quabis, R.~Dorn, M.~Eberler, O.~Glöckl,
G.~Leuchs, Appl.~Phys.~B~\textbf{72}, 109 (2001).
\bibitem{RP PRL}R.~Dorn, S.~Quabis, and G.~Leuchs, Phys.~Rev.~Lett.~\textbf{91}(23),
233901 (2003).

\bibitem{Abramowitz}M.~Abramowitz, I.~A.~Stegun, \emph{Handbook of Mathematical Functions with Formulas, Graphs, and Mathematical Tables}, (New York, Dover 1972).

\bibitem{vlpl}A.~Pukhov, J.~Plasma~Phys.\textbf{~61}, 425 (1999).
\end{thebibliography}
\end{document}